# Stability of charges in titanium compounds and charge transfer to oxygen in titanium dioxide

D Koch, P Golub, S Manzhos

Department of Mechanical Engineering, National University of Singapore, 9 Engineering Drive 1, Singapore 117576, Singapore

mpemanzh@nus.edu.sg

**Abstract**. We investigate the charge density distribution in titanium dioxide, molecular titanium complexes and a variety of periodic titanium compounds using delocalization indices and Bader charge analysis. Our results are in agreement with previous experimental and theoretical investigations on the charge stability and deviation from formal oxidation states in transition metal compounds. We present examples for practically relevant redox processes, using molecular titanium dioxide model systems, that illustrate the failure of formal oxidation states to account for some redox phenomena. We observe a pronounced charge stability on titanium for trial systems which are expected to be mainly ionic. No environment tested by us is capable to reduce the local titanium charge remainder below one electron.

## 1. Introduction

The rationalization of redox processes in terms of formally defined charge transfers between the involved reactants is, by definition, based on the change of the oxidation states of the involved species. Although the way formal oxidation states (FOS) are determined can be justified by a variety of valid arguments [1], the implications that come with them, especially with regard to redox processes, might be counter-intuitive and of little practical use. Quantum theory of atoms in molecules (QTAIM) on the other hand provides a way to quantify the electron density distribution along bonds and assign charges to the atomic centers [2]. It takes into account the continuity of the electron density, as opposed to the necessarily integer charges in FOS. Changes in the electron density distribution upon formation of new bonds can in that way be assigned to electron fluxes between different atoms, a perspective on redox reactions which aligns better with reality than the standard definition of reduction and oxidation states which dates from before quantum mechanics came into the picture but still has substantial influence on the way redox reactions are commonly understood.

Recently, a work from the group of G. Ceder revealed the oxygen redox activity in cation-disordered and Li-excess cathode materials [3]. Electron transfer on oxygen in this cases can happen due to the creation of unoccupied oxygen states into which electrons are transferred upon Li insertion, an unusual process, since in most oxide materials the oxygen-centered ligand group orbitals of the interconnected $MO_n$ polyhedra lie lower in energy than the transition metal-centered orbitals. This view on reduction and oxidation is in accordance with the formal definition of oxidation states. On the other hand, every electron transfer from an intercalant to the host lattice is going to affect the electron density distribution around all crystal atoms and will lead to changes in the QTAIM charge states of the species, which is, together with other electron density-based measures as e.g. spin density differences, commonly used to rationalize electron transfers accompanying insertion reactions [4]. According to these measures, a degree of oxygen reduction can be seen in non-defected oxides [5].

We have addressed the well-known, but nevertheless potentially problematic discrepancy between charge and oxidation states before [5, 6]. In this work, we want to underpin previously reported findings of relative charge stability in transition metal compounds with different FOS of the metal center and demonstrate the charge state changes of oxygen in oxides using simple molecular and periodic model systems. The charge states were on one hand obtained from the commonly used Bader analysis method [7]. On the other hand it is useful to exploit the concept of delocalization indices (DIs) [8], which are, although indirect, also able to provide more insight into charge transfer processes. Delocalization indices, denoted as $\delta(A,B)$, formally provide a measure for the density of the Fermi hole, created by electrons within one space domain A, that is present within another space domain B. Since a Fermi hole can be related to a certain electron and describes the decrease in the probability to find another same-spin electron nearby [9], its distribution pattern mimics the charge density distribution pattern of that electron [10]. Thus one can alternatively say that DIs measure the extent to which electrons in a given space domain A are delocalized into another space domain B, and vice versa.

The expression for the DIs can be obtained from the electron pair density $\Gamma(\vec{r_1},\vec{r_2})$, which, assuming spin-orbitals $\varphi$ of a Hartree-Fock (HF) wave function, is

$$\Gamma(\vec{r_1},\vec{r_2}) = \frac{1}{2}\sum_i \sum_j (\varphi_i^*(\vec{r_1})\varphi_i(\vec{r_1})\varphi_j^*(\vec{r_2})\varphi_j(\vec{r_2}) - \varphi_i^*(\vec{r_1})\varphi_i(\vec{r_2})\varphi_j^*(\vec{r_2})\varphi_j(\vec{r_1})). \qquad (1)$$

The first term is a simple product of individual electron densities $\rho(\vec{r_1})\rho(\vec{r_2})$, the last term is similar to the exchange term of the HF theory, correcting this product to the proper number of electron pairs $N(N-1)/2$ in the system. Formally it acts equivalently to a Fermi hole, removing the same amount of electron density [10]. Therefore its integration over two given space domains provides information about how much of the Fermi hole from the first domain delocalizes into the second domain and vice versa, that is the corresponding DI. In Fulton formulation [11]

$$\delta(A,B) = \int_A d\vec{r_1} \int_B d\vec{r_2}\, n_i^{1/2} n_j^{1/2} \varphi_i^*(\vec{r_1})\varphi_i(\vec{r_2})\varphi_j^*(\vec{r_2})\varphi_j(\vec{r_1}), \qquad (2)$$

where $n$ denotes the orbital occupancy.

Given that the chosen space domains may be associated with atoms, like quantum atoms in QTAIM [2], the value of the DIs between two such domains is naturally related to the covalent contribution to the interaction between the corresponding atoms. It was shown that with space domains taken as QTAIM atoms, the values of the DIs are related to the formal bond order [12]. Results produced by DFT in Kohn-Sham formulation are generally very similar to the results from HF calculations [13, 14]. Due to the missing electron correlation, they lead to some overestimation of the results compared to fully correlated methods. It was shown that the DIs obtained both at the DFT and HF levels of theory are sufficient to gain chemical insight [14].

Since the population of QTAIM basins – that is the one-electron density integrated over the basin – characterizes the effective atomic charges [2], both concepts, QTAIM basin populations and DIs, provide the ionic and covalent components of the chemical bond, and in combination are therefore giving a measure for its characterization.

In our investigations, we focused on titanium compounds due to their pronounced covalent character and especially on the practically relevant $TiO_2$. Titanium dioxides are promising electrode materials for Li- and post-Li ion batteries [15] and changes in their electronic structure upon metal insertion and the rationalization of the occurring redox processes are of broad interest. Titania compounds are large-bandgap semiconductors and p-doping (e.g. with Ca) can create low-lying states in the valence band, which would be expected to be occupied by upon metal intercalation [16].

We first compare DIs and Bader charge analysis on rutile $TiO_2$, followed by a discussion of the relative changes of QTAIM charge in $LiTiO_2$ and $Li_2TiO_2$ molecules as simplest model systems for lithiated titanium dioxide for which both of these stoichiometries were previously reported in experimental works [17]. Subsequently, we want to demonstrate the relative QTAIM charge stabilities in different complexes and crystals even for widely differing FOS, as well as for interstitial Ti in p-doped crystals. The aim of this work is to illustrate how direct analysis of charge density can be used to rationalize redox processes in a way uncoupled from the formal oxidation states and to illustrate this on practically important phenomena (such as lithiation) which cannot be properly described with FOS. Specifically, the lithiated $TiO_2$ molecules allow us to clearly demonstrate a degree of oxygen reduction for which FOS fail to account.

2. **Computational Details**

In order to evaluate DI's and QTAIM basin populations for rutile $TiO_2$ in section 3.1, projector-augmented wave (PAW) [18] calculations were performed as implemented in ABINIT [19], utilizing the generalized gradient approximation (GGA) with the Perdew-Burke-Ernzerhof (PBE) [20] functional. A plane-wave energy cutoff of 20 Hartree was used and a 4×4×8 k-point grid. A $3s^23p^64s^23d^2$ valence configuration was used for titanium, and a $2s^22p^4$ valence configuration was taken for oxygen. DIs were evaluated with the Dgrid-4.7 program [21]. Molecular complexes were calculated with the quantum chemical program package Gaussian 09 [22] using the Becke three-parameter Lee-Yang-Parr (B3LYP) DFT functional [23] and a *cc*-pVDZ basis set [24], relaxing all molecular structures and using the lowest-energy spin state.

The periodic structures in section 3.3 were computed with the Vienna Ab Initio Simulation Package (VASP 5.3) [25] using the PBEsol functional [26], the PAW method [18, 27] and plane-wave basis sets (500 eV cutoff energy). The used supercells were optimized using a conjugate-gradient algorithm and the valence configurations for the occurring elements were: $3s^23p^63d^24s^2$ (Ti), $4s^24p^5$ (Br), $5s^25p^5$ (I), $2s^22p^4$ (O), $2s^22p^2$ (C), $3s^23p^64s^2$ (Ca) and $2s^2$ (Be). All supercells were chosen to have dimensions of around 12x12x12 $Å^3$ and a 3x3x3 Monkhorst-Pack [28] k-point mesh was chosen in all cases. Bader charge analyses were performed using a grid-based method as implemented in the Bader v0.95a code [29]. Spin polarization was not found to be of importance for any periodic system.

## 3. Results

*3.1. Titanium dioxide charges using delocalization indices and electron transfer to oxide ions*

In rutile $TiO_2$, the titanium atom is coordinated octahedrally to six oxygen atoms, three at a distance of 1.95 Å, and other three at a distance of 1.98 Å. The corresponding DI's are equal to 0.55 and 0.48 indicating that there are approximately 3*0.55+3*0.48=3.09 electron pairs involved in six Ti-O bonds. The QTAIM population of the titanium basin is equal to 19.77, indicating transfer of 2.23 electrons from the titanium basin to the closest oxygen basins (1.11 electrons to each), which is consistent with the previous calculations: +2.23 in the titanium basin from the PBE calculations and +2.43 from the HSE06 [30] calculations [31]. Both indicators are best of all consistent with the charge state of titanium being less than +4, while the charge state of oxygen can be assumed between -1 and -2. This aligns also very well with our previous findings of Bader charges on the order of 2.5 for rutile $TiO_2$ and PBE [6]. We will further focus on Bader charge analysis, which is a reliable measure of charge distribution in molecular and bulk titanium dioxide systems, as we showed in that previous investigation [6], and is consistent with the results obtained from the DI formalism.

The attachment of one and two Li atoms to a single $TiO_2$ molecule leads to the Bader charges (expressed as atomic charges with respect to the neutral atoms) listed in Table 1. The obtained bent geometries for $LiTiO_2$ (tetragon-like arrangement of the four atoms) and $Li_2TiO_2$ (pentagon-like) correspond to the predicted ground-state structures, while the linear mono- and dilithiated $TiO_2$ are saddle points of the potential energy surface (the spin states of the mono- and dilithiated compounds are doublet and triplet, respectively, in both geometric arrangements). The latter are included since linear O-Ti-O arrangements are common structural motifs in crystalline titania phases. We use the mono- and dilithiated molecules as simplest models for the local electron transfer in bulk $TiO_2$ (e.g. rutile or anatase which both include linear and bent O-Ti-O units). Lithiation of bulk $TiO_2$ creates localized in-gap states [32], which is why we deem a single molecule a meaningful case study for local electron transfers in lithiated, solid $TiO_2$. However, the more pronounced electron delocalization in periodic systems does not allow for a quantitative transferability of the following results to the solid-state case, but is supposed to provide mechanistic insight and values at the limit of fully localized charges. Representations of all four molecules together with charge density isosurface plots are shown in Fig. 1 (visualization was done with the program VESTA [33]).

**Table 1.** Bader charges of atoms in $TiO_2$, $LiTiO_2$ and $Li_2TiO_2$ in linear and bent configuration calculated with B3LYP/*cc*-pVDZ.

| | Bader charge [/*e*/] | | | |
|---|---|---|---|---|
| **Compound** | **Ti** | **O1** | **O2** | **Li** |
| $TiO_2$ (bent) | +2.11 | -1.06 | -1.06 | - |
| $LiTiO_2$ (bent) | +1.66 | -1.33 | -1.32 | +0.98 |
| $Li_2TiO_2$ (bent) | +1.74 | -1.40 | -1.42 | +0.54 |
| $TiO_2$ (linear) | +2.43 | -1.22 | -1.22 | - |
| $LiTiO_2$ (linear) | +1.97 | -1.26 | -1.70 | +0.98 |
| $Li_2TiO_2$ (linear) | +1.53 | -1.75 | -1.74 | +0.97 |

As can be seen from Table 1, the charges on oxygen upon attachment of Li, which is directly bound to it, change by -0.27 and -0.07 $|e|$ from non- to monolithiated and mono- to dilithiated bent $TiO_2$ respectively, while the charge on Ti decreases by 0.45 $|e|$ and then increases by 0.08 $|e|$ due to existence of a non-nuclear attractor between the two Li ions reducing the charge donation to the $TiO_2$ fragment (see inset of Fig. 1). The charge in the additional basin of the non-nuclear attractor was divided between the two Li centers and was used to assign a Bader charge to both ions. The delocalization of charge over the whole molecule leads to a transfer of electron density to Ti and each O which are comparable in magnitude.

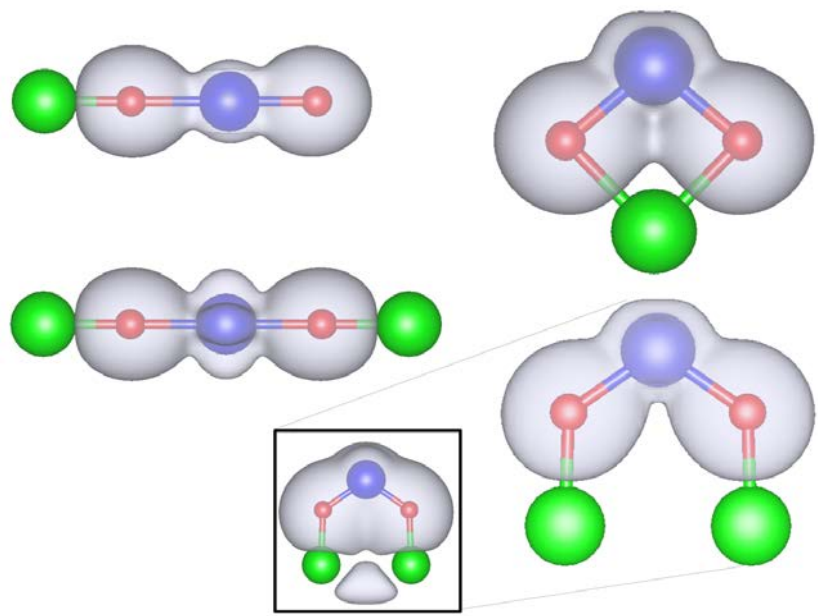

**Figure 1.** Geometries of four optimized lithiated $TiO_2$ molecules. From top left to bottom right: $LiTiO_2$ (linear), $LiTiO_2$ (bent), $Li_2TiO_2$ (linear), $Li_2TiO_2$ (bent). Isosurface plots of the charge densities in every molecule (isosurface value: 0.03 a.u.) are shown in grey. Ti atoms are represented as blue, oxygen as red and lithium as green coloured spheres. The inset shows an isosurface plot with a lower isovalue (0.006 a.u.) for bent $Li_2TiO_2$ in order to visualize the non-nuclear attractor between the two Li ions.

In the linear lithium titanium dioxides, both attached lithium atoms are predicted to lose most of their valence density, transferring a charge of -0.90 $|e|$ on Ti and -0.53 $|e|$ on each oxygen. In other words, a 'whole' electron is transferred on both oxygen atoms and one electron on Ti upon dilithiation, based on the local change of electron density. Analogous trends, which can be traced back to the covalency of the Ti-O bonds, have been reported previously by Albaret *et al.* connecting charge transfers in $Ti_nO_m^x$ ($n = 1,2,3$; $m-n=0,1$; $x=-1,0,1$) to changes in orbital hybridization [34]. From a charge-density point of view, the oxygen ions are being 'reduced'. This on one hand is not well handled by the FOS formalism and on the other hand this shows that oxygen redox activity in battery electrode materials, which has been rationalized in recent literature by invoking O defects, must be present in undefected oxides.

*3.2.    Charge stability in molecular titanium complexes*
In a recent work, Wolczanski presented a way to estimate the relative charge states of transition metal ions in carbonyl complexes, utilizing the weakening of the C-O bond upon reduction of the carbonyl ligands [35]. Using the average of reported C-O stretch frequencies in different Fe complexes and assuming a stable +2 $|e|$ charge state on the iron center, he could deduce the charge state of a variety of transition metal carbonyl complexes by comparison of the C-O stretch frequencies, leading to a charge state estimation of +3.2 $|e|$ for Ti in $[Ti(CO)_6]^{2-}$. The results of this charge distribution via reporters (CDVR) method were backed up by Mulliken and Natural Bond Order analyses. We decided to verify if Bader charge analyses are in accordance with the results obtained from the formalism in ref. 35.

Fig. 2 shows the average C-O stretch frequencies (without considering degeneracies) versus the assumed CO ligand charges (with assumed fixed +2 charge state of Fe) for $[Fe(CO)_4]^{2-}$ ($q_{CO} = -1$ $|e|$), $Fe(CO)_5$ ($q_{CO} = -0.4$ $|e|$) and $[Fe(CO)_6]^{2+}$ ($q_{CO} = 0$ $|e|$) calculated with DFT. Although the experimentally observed frequencies could not be matched accurately (overestimation on the order of 100 cm$^{-1}$ (7 %)), which might be caused by non-vacuum frequency values used in the data basis of ref. 35, a linear trend can be observed. Using the average C-O stretch in $[Ti(CO)_6]^{2-}$ at 1875 cm$^{-1}$ and comparing with the linear trend line in Fig. 2 (red), the computed charge state on Ti is +3.5 $|e|$. The charge state deduced from the CDVR method was given as +3.2 $|e|$, but the actual interpolated value obtained by Wolczanski was 3.7-3.8 $|e|$. It was modified due to consistency considerations with regard to the other investigated complexes. With this initially obtained value however, a reasonable level of agreement can be observed.

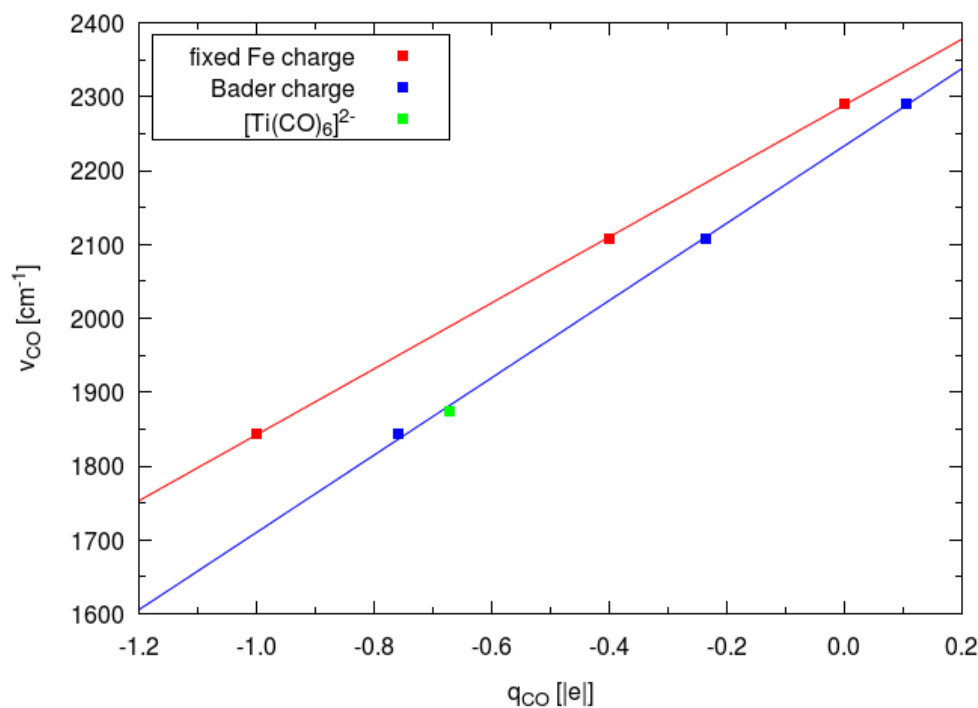

**Figure 2.** Computed average vibrational frequencies of the C-O stretch in $[Fe(CO)_4]^{2-}$, $Fe(CO)_5$ and $[Fe(CO)_6]^{2+}$ against assumed fixed CO ligand charges (red color, $q = -1.0$ $|e|$, $-0.4$ $|e|$, $0.0$ $|e|$) or calculated Bader charges from DFT (blue). The (v, $q_{Bader}$) pair for $[Ti(CO)_6]^{2-}$ is indicated in green and is in reasonable agreement with the blue linear trend line.

Up to this point, the ligand charges were based on an assumed charge state of Fe which has been fixed to +2. Bader charge analyses of the three iron complexes indicate charges of +1.02 $|e|$ in $[Fe(CO)_4]^{2-}$, +1.18 $|e|$ in $Fe(CO)_5$ and +1.37 $|e|$ in $[Fe(CO)_6]^{2+}$ on Fe, while the titanium charge state is +2.03 $|e|$ in $[Ti(CO)_6]^{2-}$. Plotting again symmetric C-O stretch frequencies, this time versus CO Bader charges for all four complexes, the linear relationship in Fig. 2 (blue) was obtained. It can be deduced that CO as a charge reporter ligand is indeed capable to give an estimate of the relative charge state, but the QTAIM Bader charge on Fe in the reference complexes is closer to +1 than +2, which is the reason for the overestimation of the Ti charge state compared to theory. Further, the charges on Fe are not expected to be perfectly constant, which contributes to a small further mismatch with the DFT results.

Charge states of +2.03 $|e|$, +2.11 $|e|$ and +2.43 $|e|$ for the vastly different molecular complexes $[Ti(CO)_6]^{2-}$ as well as bent and linear $TiO_2$, respectively (cf. Table 1), indicate that a substantial charge remainder on the Ti center is favored. This is not surprising, since density delocalization and the following decrease in kinetic energy is the main driving force for bond formation [36]. The requirement for a ligand to abstract all (most) of the Ti valence density to observe complete iconicity of Ti in the context of the QTAIM analysis, does not seem to be met by common ligands in molecular Ti complexes as indicated in Table 2. Moreover, a stable charge remainder of at least one electron remains in these complexes.

**Table 2.** Bader charges on titanium for several molecular complexes and complex ions calculated with B3LYP/*cc*-pVDZ, unless indicated differently.

| Compound | Bader charge (Ti) [/$e$/] | Compound | Bader charge (Ti) [/$e$/] |
|---|---|---|---|
| $TiF_4$ | +2.46 | $Ti(CN)_4$ | +2.92 |
| $TiCl_4$ | +2.72 | $[Ti(CO)_6]^{2-}$ | +2.03 |
| $TiBr_4$ | +2.66 (+2.56[b]) | $Ti(TCNE)_2$ (tetragonal)[a] | +1.78 |
| $[TiF_6]^{2-}$ | +2.75 | $Ti(TCNE)_2$ (tetrahedral)[a] | +2.30 |

[a]TCNE: tetracyanoethylene; used as bidentate ligand here.
[b]Value obtained with PBE/*cc*-pVDZ.

3.3. *Solid titanium halogenides and titanium insertion in p-doped crystals: incomplete charge transfer*
Naturally the question occurs whether a charge delocalization in periodic systems can contribute to a further lowering of the local charge density around Ti atoms towards a more ionic situation. Previous investigations of our group [6] showed that Bader charges obtained from calculations with B3LYP and PBE do not differ substantially, so we assumed Bader charges from molecular densities obtained with B3LYP and from periodic densities with PBEsol to be equally reliable for other titanium compounds.

In Table 3 the Bader charges on titanium in the crystalline titanium halogenides $TiBr_4$ and $TiI_4$ (space group: $P2_1/a3$ in both cases) are listed. As was observed for $TiO_2$ in a previous work [6], the charge on Ti increases from a molecular to a periodic system, since more delocalized background charge contributes to the Ti basin. This seems to be true for the halogenides as well, as the direct comparison of the Bader charge obtained with PBE for $TiBr_4$ in Table 2 and its Bader charge given in Table 3 (from PBEsol) shows. However, since the Bader charges reported in Table 3 were obtained using the GGA-type PBEsol functional in contrast to the hybrid B3LYP functional employed for the molecular systems, a direct comparison between Tables 2 and 3 is eventually not meaningful, due to the more pronounced self-interaction in GGA functionals compared to hybrid ones.

**Table 3.** Bader charges on titanium in crystalline halogenides and as inserted species in several crystalline host structures (Bader charges for interstitial Ti).

| Compound | Bader charge (Ti) [/$e$/] | Compound | Bader charge (Ti) [/$e$/] |
|---|---|---|---|
| $TiBr_4$ | +1.70 | Ti in Be-doped diamond-C ($TiBe_3C_{213}$) | +1.46 |
| $TiI_4$ | +1.50 | Ti in Ca-doped rutile-$TiO_2$ ($Ca_4Ti_{29}O_{64}$) | +2.09 |
| Ti in I ($TiI_{54}$) | +1.39 | | |

Another approach that was tried to nudge a further reduction of electron density on Ti was the insertion of Ti as interstitial atom in crystalline compounds whose Fermi levels are expected to be substantially lower than the newly created interstitial Ti states and having unoccupied states around the Fermi level. We chose C-diamond (space group: Fd-3m) as insulator p-doped with substitutional Be, rutile-$TiO_2$ (space group: $P4_2$/mnm) as semiconductor p-doped with substitutional Ca and metallic iodine (space group: Im-3m). Titanium was expected to donate its valence electrons to the low-lying states around the Fermi level on one hand and reduce strain energy of the host lattice by reducing its radius which could be accomplished by further ionization. As can be seen in Table 3, this is not the case. Higher-lying Ti levels are still occupied and there is a substantial electron density remainder on the interstitial Ti in all cases. Relative charge stability and charge remainders on transition metal centers in compounds commonly classified as ionic were reported in the past [37], and the same trend can be observed, besides the molecular systems in Table 2, also for the two types of environment (cation in titanium halogenide vs. interstitial solute in doped semiconductor) of Ti in the periodic systems presented in Table 3.

## 4. Conclusions

We provided, using QTAIM methods, additional insight into the charge density distributions in common titanium compounds, focusing on titanium dioxide, halogenides and a hexacarbonyl complex. Of course the Bader charge analyses do not technically contradict the definition of formal oxidation states and in this framework they could be simplistically explained by a pronounced covalent character of the investigated compounds. Nevertheless the observed charge stability in a variety of titanium compounds and the Ti atom's apparent resistance to strong electron deficiency leads to counter-intuitive implications when it comes to charge density fluxes in redox processes, such as the lithiation of $TiO_2$, a reaction occurring in titania electrodes of Li-ion batteries. We could theoretically reproduce experimentally observed trends [35] with regard to the charge state in titanium hexacarbonyl and show that Bader charge analysis results can in principle be backed up with concrete experimental evidence and are not just a computational quirk.

We previously reported that the QTAIM charge in the solid state is increased compared to the molecular setup due to a higher delocalized background charge in $TiO_2$ [6]. We obtained a similar result for titanium halogenides and tested the capability of p-doped periodic systems to ionize interstitially inserted Ti atoms, which is poor.

We furthermore presented the use of delocalization indices to evaluate the charge state of titanium in rutile $TiO_2$, supporting the claimed reliability of Bader charges by a more sophisticated systematic theoretical analysis method and providing insight into the peculiarities of the electron density distribution in this compound. From the results of the limited number of trial systems utilized, there seems to be a tendency in titanium compounds to keep at least one electron localized around titanium, although Bader charges of +4 $|e|$ and more are not prohibitive and are possible in general (e.g. $q = +5.8$ $|e|$ on S in $SF_6$).

## Acknowledgements

This work was supported by the Ministry of Education of Singapore (grant no. MOE2015-T2-1-011).

## References

[1] Karen P, McArdle P and Takats J 2016 Comprehensive definition of oxidation state (IUPAC Recommendations 2016) *Pure Appl. Chem.* **88** 831
    Resta R 2008 Charge states in transition *Nature* **453** 735
[2] Bader R F W 1985 Atoms in molecules *Acc. Chem. Res.* **18** 9
[3] Seo D-H, Lee J, Urban A, Malik R, Kang S and Ceder G 2016 The structural and chemical origin of the oxygen redox activity in layered and cation-disordered Li-excess cathode materials *Nat. Chem.* **8** 692
[4] Yaresko A N, Antonov V N, Eschrig H, Thalmeier P and Fulde P 2000 Electronic structure and exchange coupling in α'-$NaV_2O_5$ *Phys. Rev. B* **62** 15538
    Arrouvel C, Parker S C and Islam M S 2009 Lithium insertion and transport in the $TiO_2$−B anode material: A computational study *Chem. Mater.* **21** 4778
[5] Kulish V and Manzhos S 2017 Comparison of Li, Na, Mg and Al-ion insertion in vanadium pentoxides and vanadium dioxides *RSC Adv.* **7** 18643
[6] Koch D and Manzhos S 2017 On the charge state of titanium in titanium dioxide *J. Phys. Chem. Lett.* **8** 1593
[7] Bader R F W 1998 Atoms in molecules. In *Encyclopedia of Computational Chemistry* (Hoboken: John Wiley & Sons, Inc) p 64
[8] Bader R F W and Stephens M E 1975 Spatial localization of the electronic pair and number distributions in molecules *J. Am. Chem. Soc.* **97** 7391
[9] McWeeny R 1960 Some recent advances in density matrix theory *Rev. Mod. Phys.* **32** 335
[10] Bader R F W, Streitwieser A, Neuhaus A, Laidig K E and Speers P 1996 Electron delocalization and Fermi hole *J. Am. Chem. Soc.* **118** 4959
[11] Fulton R L and Mixon S T 1993 Comparison of covalent bond indices and sharing indices *J. Phys. Chem* **97** 7530
[12] Fradera X, Austen M A and Bader R F W 1999 The Lewis model and beyond *J. Phys. Chem.* **103** 304
[13] Kar T, Angyan J G and Sannigrahi A B 2000 Comparison of ab initio Hartree-Fock and Kohn-Sham orbitals in the calculation of atomic charges, bond index, and valence *J. Phys. Chem. A* **104** 9953
[14] Poater J, Sola M, Duran M and Fradera X 2002 The calculation of electron localization and

| | delocalization indices at the Hartree-Fock, density functional and post-Hartree-Fock levels of theory *Theor. Chem. Acc.* **107** 362 |
|---|---|
| [15] | Liu J, Wang J, Ku Z, Wang H, Chen S, Zhang L, Lin J and Shen Z X 2016 Aqueous rechargeable alkaline $Co_xNi_{2-x}S_2/TiO_2$ Battery *ACS Nano* **10** 1007 |
| | Oh S-M, Hwang J-Y, Yoon C S, Lu J, Amine K, Belharouak I and Sun Y-K 2014 High electrochemical performances of microsphere $C-TiO_2$ anode for sodium-ion battery *ACS Appl. Mater. Interfaces* **6** 11295 |
| [16] | Lueder J, Cheow M H and Manzhos S 2017 Understanding doping strategies in the design of organic electrode materials for Li and Na ion batteries: an electronic structure perspective *Phys. Chem. Chem.Phys.* **19** 13195 |
| | Legrain F and Manzhos S 2015 Aluminum doping improves the energetics of lithium, sodium, and magnesium storage in silicon: a first-principles study *J. Power Sources* **274** 65 |
| | Lueder J, Legrain F, Chen Y and Manzhos S 2017 Doping of active electrode materials for electrochemical batteries: an electronic structure perspective *MRS Commun.* **7** 523 |
| [17] | Borghols W J H, Lutzenkirchen-Hecht D, Haake U, van Eck E R H, Mulder F M and Wagemaker M 2009 The electronic structure and ionic diffusion of nanoscale $LiTiO_2$ anatase *Phys. Chem. Chem. Phys.* **11** 5742 |
| | Huang Y F, Tang H, Li J, Hu J, Zhu X Q, Zhang W J, Shen L and Qing L X 2012 Preparation of Lithium Orthosilicate Ceramic Pebbles by Molten Spray Method Process *Adv. Mater.* **412** 111 |
| [18] | Bloechl P E 1994 Projector augmented-wave method *Phys. Rev. B: Condens. Matter. Mater. Phys.* **50** 17953 |
| [19] | Gonze X, Beuken J-M, Caracas R, Detraux F, Fuchs M, Rignanese G-M, Sindic L, Verstraete M, Zerah G, Jollet F, Torrent M, Roy A, Mikami M, Ghosez P, Raty J-Y and Allan D C 2002 First-principles computation of material properties: The ABINIT software project *Comput. Mater. Sci* **25** 478 |
| | Gonze X, Amadon B, Angade P-M, Beuken J-M, Bottin F, Boulanger P, Bruneval F, Caliste D, Caracas R, Cote M, Deutsch T, Genovese L, Ghosez P, Giantomassi M, Goedecker S, Hamann D R, Hermet P, Jollet F, Jomard G, Leroux S, Mancini M, Mazevet S, Oliveira M J T, Onida G, Pouillon Y, Rangel T, Rignanese G-M, Sangalli D, Shaltaf R, Torrent M, Verstraete M J, Zerah G and Zwanziger J W 2009 ABINIT: First-principles approach to material and nanosystem properties *Comput. Phys. Commun.* **180** 2582 |
| [20] | Perdew J P, Burke K and Ernzerhof M 1996 Generalized gradient approximation made simple *Phys. Rev. Lett.* **77** 3865 |
| [21] | Kohout M 2012 *DGrid-4.7* (Radebeul) |
| [22] | Frisch M J, Trucks G W, Schlegel H B, Scuseria G E, Robb M A, Cheeseman J R, Scalmani G, Barone V, Petersson G A, Nakatsuji H, Li X, Caricato M, Marenich A, Bloino J, Janesko B G, Gomperts R, Mennucci B, Hratchian H P, Ortiz J V, Izmaylov A F, Sonnenberg J L, Williams-Young D, Ding F, Lipparini F, Egidi F, Goings J, Peng B, Petrone A, Henderson T, Ranasinghe D, Zakrzewski V G, Gao J, Rega N, Zheng G, Liang W, Hada M, Ehara M, Toyota K, Fukuda R, Hasegawa J, Ishida M, Nakajima T, Honda Y, Kitao O, Nakai H, Vreven T, Throssell K, Montgomery Jr V, Peralta J E, Ogliaro F, Bearpark M, Heyd J J, Brothers E, Kudin K N, Staroverov V N, Keith T, Kobayashi R, Normand J, Raghavachari K, Rendell A, Burant J C, Iyengar S S, Tomasi J, Cossi M, Millam J M, Klene M, Adamo C, R. Cammi, Ochterski J W, Martin R L, Morokuma K, Farkas O, Foresman J B and Fox D J 2016 *Gaussian 09* Revision D.01 (Wallingford CT: Gaussian, Inc.) |
| [23] | Lee C, Yang W and Parr R G 1988 Development of the Colle-Salvetti correlation-energy formula into a functional of the electron density *Phys. Rev. B* **37** 785 |
| | Becke A D 1993 Density-functional thermochemistry. III. The role of exact exchange *J. Chem. Phys.* **98** 5648 |
| | Devlin F J, Finley J W, Stephens P J and Frisch M J 1995 Ab initio calculation of vibrational absorption and circular dichroism spectra using density functional force fields: A comparison of local, nonlocal, and hybrid density functionals *J. Phys. Chem.* **99** 16883 |
| [24] | Balabanov N B and Peterson K A 2005 Systematically convergent basis sets for transition metals. I. All-electron correlation consistent basis sets for the 3d elements Sc–Zn *J. Chem. Phys.* **123** 64107 |
| | Balabanov N B and Peterson K A 2006 Basis set limit electronic excitation energies, ionization |

potentials, and electron affinities for the 3d transition metal atoms: coupled cluster and multireference methods *J. Chem. Phys.* **125** 74110.

Dunning Jr T H 1989 Gaussian basis sets for use in correlated molecular calculations. I. The atoms boron through neon and hydrogen *J. Chem. Phys.* **90** 1007

[25] Kresse G and Furthmüller J 1996 Efficient iterative schemes for ab initio total-energy calculations using a plane-wave basis set *Phys. Rev. B* **54** 11169

[26] Perdew J P, Ruzsinszky A, Csonka G I, Vydrov O A, Scuseria G E, Constantin L A, Zhou X and Burke K 2008 Restoring the density-gradient expansion for exchange in solids and surfaces *Phys. Rev. Lett.* **100** 136406

[27] Kresse G and Joubert D 1999 From ultrasoft pseudopotentials to the projector augmented-wave method *Phys. Rev. B* **59** 1758

[28] Monkhorst H J and Pack J D 1976 Special points for Brillouin-zone integrations *Phys. Rev. B* **13** 5188

[29] Tang W, Sanville E and Henkelman G 2009 A Grid-based Bader analysis algorithm without lattice bias *J.Phys. Condens. Matter* **21** 84204

Henkelman G, Arnaldsson A and Jónsson H 2006 A Fast and robust algorithm for Bader decomposition of charge density *Comput. Mater. Sci.* **36** 354

Sanville E, Kenny S D, Smith R and Henkelman G 2007 Improved grid-based algorithm for Bader charge allocation *J. Comput. Chem.* **28** 899

Yu M, Trinkle D R 2011 Accurate and efficient algorithm for Bader charge integration *J. Chem. Phys.* **134** 64111

[30] Krukau A V, Vydrov O A, Izmaylov A F and Scuseria G E 2006 Influence of the exchange screening parameter on the performance of screened hybrid functionals *J. Chem. Phys.* **125** 224106

[31] Vasquez G C, Karazhanov S Zh, Maestre D, Cremades A, Piqueras J and Foss S E 2016 Oxygen vacancy related distortions in rutile $TiO_2$ nanoparticles: A combined experimental and theoretical study *Phys.Rev. B* **94** 235209

[32] Legrain F, Malyi O I and Manzhos S 2015 Insertion energetics of lithium, sodium, and magnesium in crystalline and amorphous titanium dioxide: A comparative first-principles study *J. Power Sources* **278** 197

[33] Momma K and Izumi F 2011 VESTA 3 for three-dimensional visualization of crystal, volumetric and morphology data *J. Appl. Crystallogr.* **44** 1272

[34] Albaret T, Finocchi F and Noguera C 1999 First principles simulations of titanium oxide clusters and surfaces *Faraday Discuss.* **114** 285

[35] Wolczanski P 2017 Flipping the oxidation state formalism: charge distribution in organometallic complexes as reported by carbon monoxide *Organometallics* **36** 622

[36] Schmidt M W, Ivanic J and Ruedenberg K 2014 Covalent bonds are created by the drive of electron waves to lower their kinetic energy through expansion *J. Chem. Phys.* **140** 204104

[37] Raebiger H, Lany S and Zunger A 2008 Charge self-regulation upon changing the oxidation state of transition metals in insulators *Nature* **453** 763

Christensen A and Carter E A 2000 First-principles characterization of a heteroceramic interface: $ZrO_2(001)$ deposited on an α-$Al_2O_3(1102)$ substrate *Phys. Rev. B* **62** 16968